# DESIGN AND PERFORMANCE OF "R-FLEX", A FLEXURE-BASED FIBER POSITIONING ROBOT FOR SPECTROSCOPIC COSMOLOGY

**Nicholas R. Wenner, Joseph H. Silber, David J. Schlegel**
Lawrence Berkeley National Laboratory
Berkeley, CA, USA

## INTRODUCTION

Cosmology and astrophysics are presently experiencing an amazing revolution in the quantity and precision of data on the large scale structure of the universe. At the forefront of the field is the Dark Energy Spectroscopic Instrument (DESI), a telescope instrument with 5000 robotically actuated optical fibers [1]. These fibers deliver the light from faint galaxies, billions of light years away, to high resolution spectrographs positioned off-telescope. To date, DESI has collected over 56 million spectra.

The key enabling technology for DESI was the development of miniature, high precision robotic positioners, which could be mounted in a close-packed array as tightly as 10.4 mm center-to-center pitch. The next generation of massively parallel fiber-fed instruments will require fiber robots about 2.5-3x smaller by area [2].

In the present work we describe design, fabrication, and testing of a novel flexure mechanism ("R-FLEX", FIGURE 1) which provides precision radial motion for fiber robots that can be mounted at 6.2 mm pitch.

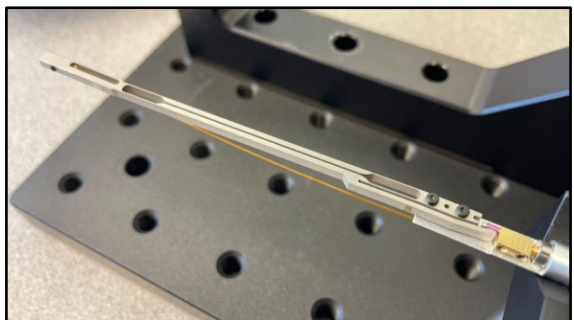

FIGURE 1. R-FLEX fiber robot.

The design must deliver repeatable, precision motion (≤ 5 µm) over a relatively large travel range (4 mm), that extends outside each unit's quite small overall packaging envelope (ø5.8 mm). It must be amenable to mass production (~30k units), have < ~30 µm parasitic (out of plane) motion, < ~0.3° angular misalignment, and ~1 million cycle lifetime operating in a mountaintop telescope environment.

## DESIGN

It has long been recognized that polar, backlash-free kinematics would be desirable for fiber robots, but prior designs have tended toward either relatively large mechanisms or relatively complex flexures [3][4]. The R-FLEX design presented here achieves a slim package, low part-count, and natural backlash rejection.

R-FLEX converts small rotational motions at the flexure base into large linear motions at the output (FIGURE 2). A roller cam is driven in precision increments by a ø4 mm brushless DC gearmotor, with a reduction ratio of > ~100:1. The roller pushes a 19 mm arm, putting a rotational moment on a 60 mm parallel-arm flexure. At either end of the arms are pairs of symmetric thin leaf flexures. The result is a very flat motion with nearly zero angular tilt.

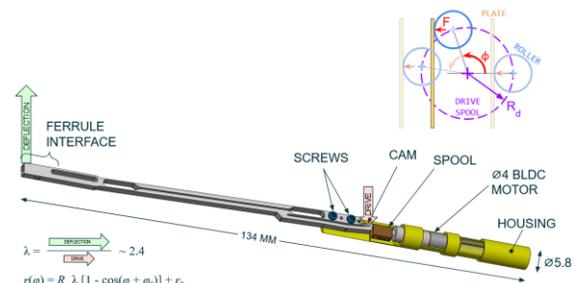

FIGURE 2. Design of R-FLEX assembly.

We developed the design in four phases: (1) Developed kinematic and stress calculations in a design spreadsheet. (2) Converted calculations to a Python script and ran 100,000 iterations of design parameters. (3) Selected the best design and performed finite element analysis to confirm stress and deflection. (4) Built & tested prototypes to confirm FEA.

### Material Selection and Fabrication

We studied several candidate flexure materials, in particular Ti 6Al-4V, 17-7PH (TH1050), 7075-T6, and BeCu C17200 (TH04). We selected the titanium for (1) its high ratio of fatigue strength to elastic modulus, and (2) no need for heat treatment, which would add fabrication challenges for our thin-walled, complex, precision geometry.

The flexure was fabricated using wire EDM, which afforded tight dimensional tolerances and efficient production. Given the 2D profile of the flexure, multiple flexures could be fabricated at once via stacking. Remaining parts were procured commercially or fabricated with traditional CNC machining, which also lends itself to large-scale manufacturing.

**PERFORMANCE**

Tip, tilt, and focus were measured using an optical CMM. Metrics related to r-y position were measured using a custom test station employing optical centroiding of back-lit fibers. CMM measurements were carried out for 5 flexure positions across the travel range. Centroiding measurements were carried out for "blind" and "correction" moves. Blind moves represent large initial moves that are completed at high speeds. Correction moves represent subsequent and more precise motions that occur over shorter distances and at lower speeds to maximize final positional accuracy. Blind move measurements were carried out for 1000 random targets throughout the range, whereas correction move measurements were carried out in 1 deg increments in motor angle for 10 complete cycles across the range (n = 3360). Additionally, the natural frequency was measured by inspection of high-speed video.

*TABLE 1. Representative comparisons of predicted and measured values.*

| ITEM | PREDICTED | MEASURED |
|---|---|---|
| r-y Error | - | 4.2 µm rms |
| Natural Frequency | 35.3 Hz | 37.3 Hz |
| r Angular Error | 0.029 deg | 0.020 deg |
| y Angular Error | 0.140 deg | 0.086 deg |
| Defocus | 54 µm | 50 µm |
| Travel Range | 3.799 mm | 4.171 mm |

A calibration function for position (r) as a function of motor angle (φ) was fit to measured data, with representative values shown below:

$$r(\varphi) = R_d \lambda [1 - \cos(\varphi + \varphi_o)] + r_o$$

$R_d \lambda = 2.075$ mm
$\varphi_o = 1.004$ deg
$r_o = 0.005$ mm

**Discussion**

Measured performance exceeded predicted performance for r angular error, y angular error, and defocus. Natural frequency and travel range were within 10% of predicted values.

The assembly met r-y position performance of ≤5 µm rms at high and low speeds for the extension direction of the flexure and at high speed for the retraction direction. Larger positional errors were seen for slow movements in the retraction direction, potentially related to asymmetries in static friction. As such, correction moves may be best accomplished in the extension direction.

**CONCLUSIONS**

R-FLEX is a flexure-based radial stage for precision robotic motion of optical fibers. Performance of prototypes aligned closely with predicted design values. The design is a promising candidate for delivering better than 5 µm positioning accuracy over large 4 mm travel ranges in a tight ø5.8 mm package that is suitable for mass deployment on the next generation of fiber-fed spectroscopic telescope instruments.

**REFERENCES**

[1] Silber, J., et al. (2023). The Robotic Multiobject Focal Plane System of the Dark Energy Spectroscopic Instrument. The Astronomical Journal, 165(9). https://doi.org/10.3847/1538-3881/ac9ab1
[2] Silber, J., et al. (2022). 25,000 optical fiber positioning robots for next-generation cosmology. 37th Annual Meeting of ASPE. https://doi.org/10.48550/arXiv.2212.07908
[3] Schlegel, D., & Ghiorso, W. (2008). LBNL fiber positioners for wide-field spectroscopy. Proceedings of SPIE, 7018. https://doi.org/10.1117/12.801673
[4] Silber, J., et al. (2012). Design and performance of an R-θ fiber positioner for the BigBOSS instrument. Proceedings of SPIE, 8450, 845038. https://doi.org/10.1117/12.926457